# How Will the Internet of Things Enable Augmented Personalized Health?

Amit Sheth, Utkarshani Jaimini, Hong Yung Yip

Kno.e.sis Center - Wright State University

Introduction

**Internet-of-Things** (IoT) refers to network-enabled technologies including mobile and wearable devices, sensing and actuation devices that interact with each other and communicate over the internet. IoT is profoundly redefining the way we create, consume, and share information. Health aficionados and citizens are increasingly using IoT technologies to track their sleep, food intake, activity, body vital signals, and other physiological observations. This is complemented by IoT systems that continuously collect data from the environment and inside the living quarters that can affect a human health. Together, these have created an opportunity for a new generation of healthcare solutions.

The paradigm shift of reactive medicine to proactive and preventive medicine is primarily motivated by economic imperatives such as the rising cost of healthcare, as well as continued improvements to the quality of life and longevity. According to the Center for Medicare and Medicaid Services (CMS), the cost of medical care in the US in 2016 reached up to 3.6 trillion dollars per year and is expected to increase up to $5.5 trillion by 2025 [http://bit.ly/2ATjX3l]. On the other hand, the global smart healthcare industry is expected to reach $169.30 billion by 2020. It is projected that by 2019, 87% of the healthcare organizations in the US will have adopted IoT technology [http://bit.ly/2eP8wQo], of which 73% will be used to reduce cost, and 64% will be put to monitor patients.

IoT data itself is not adequate to understand an individual's health, and associated aspects of wellbeing and fitness; it is usually necessary to look at that individual's clinical record and behavioral information, as well as social and environmental information affecting that individual. Interpreting how well a patient is doing also requires looking at his adherence to respective health objectives, application of relevant clinical knowledge and the desired outcomes, such as the patient's preference for quality of life versus longevity and expert knowledge.

**Augmented Personalized Healthcare (APH)** is a vision [2, http://bit.ly/k-APH] to exploit the extensive variety of relevant data and medical knowledge using Artificial Intelligence (AI) techniques to extend and enhance human health and well-being. It anticipates the use of physical, cyber, and social data obtained from wearables and IoT devices, clinical information including Electronic Medical Records (EMRs), mobile applications that support targeted interactions and engagement with the patients, web-based information including web services (e.g. those providing health relevant data such as air quality and allergens), social media (e.g. post by patients with similar concerns and conditions), and extensive knowledge of clinical practice and medicine. The data consists of a set of signals collected at personal, public, and population level as well as knowledge that affects human health. *Augmentation* refers to aggregating this data and converting into actionable information that can improve health-related outcome due to better and more timely decisions. This embodiment of APH is an entirely new approach to human health compared to the current episodic system of periodic care primarily centered around health care establishments (such as clinics, hospitals, and labs). APH involves continuous monitoring, engagement, and health/ management, where rather than treating a patient with a disease, the focus shifts to involving the patient in preventing disease, predicting possible adverse outcomes and preventing them through proactive measures, and trying to keep citizens healthy and fit with lifestyle changes. Rather than only focusing on the management of chronic conditions, APH proposes a holistic approach for improving the overall quality of life.

**Patient-Generated Health Data** (PGHD) is the heart of APH. It is primarily generated by IoT devices and it captures the digital footprint representative of patients' health across time with finer details that are distinct from the data generated in clinical settings such as EMRs and Personal Health Records (PHR). The two main IoT categories for patient health monitoring are wearable sensors and environmental sensors. Wearable sensors are portable sensors that patients' wear the most, if not all of the time. These close vicinity wearable sensors monitor patient's physiological markers, such as heart rate, breathing rate, and blood pressure. They are designed to integrate into patients' daily routine to enable passive and continuous sensing and monitoring for timely interventions. Environmental sensors, on the other hand, are sensors that collect environmental data in the living environment of the citizens. These sensors are normally not portable, but admittedly provide critical information for health management. For example, weather data such as humidity, pollen index, and air quality, are important for asthma management. However, this data provides a coarse measure in a population level and do not take account of differences of every individual patient. To mitigate that, different sensors can be utilized as a complementarity. For example, Foobot [https://foobot.io/] monitors indoor air quality and reflects a closer overview of a patient's environment. Hence, these sensors enable personalization and allow both physicians and patients to monitor asthma at a finer level.

**Stages of Technology Enabled Health Augmentation**

In this section, we review various stages of augmented health management strategies using APH technology.

**(1) Self Monitoring:** Currently, doctors see patients infrequently or as needed for new conditions, or for those with chronic conditions and disease, a patient is seen at a well-defined time intervals (e.g., monthly, quarterly) depending on the established medical protocols and severity of the medical condition. A doctor's understanding of a patient's condition often comes primarily from patient's self description (self-reporting) and the observations gathered from the patient in person. This has limitations that not all significant events/ issues may be recalled at the time of the meeting, and the exact timing, location, and reasons of the triggering event may not be available. With continuous monitoring using IoT-enabled sensor devices, wearables, and periodically administered the contextually-relevant questionnaire, we can better capture relevant aspects of a patient's surroundings, diet, activities, and other factors related to health. All of these aspects when analyzed can help to determine possible and precise causes of patient's condition and well-being. PGHD plays an important role in supplementing existing clinical data and filling in gaps in information on an ongoing basis, thus generating a more comprehensive picture of ongoing patient health [http://bit.ly/1WIzqb1].

**(2) Self Appraisal:** Self appraisal is the ability of the patient to evaluate the relevance of a variety of data and observations within the context of the patient's health objectives or concern. Wearable devices are used to keep track of patient's day to day activities. However, there is a big gap between simply having the access to relevant data implied by self-monitoring, and interpretation of the data within the context of health objective or concern. A patient would be interested in understanding if he/she is keeping up with progress towards the health goals? Consider use of a Fitbit [11, http://bit.ly/1qYlDLV] to measure the number of steps every day and the quality of sleep. Is this data helping the patient in attaining their desired objective or they need to do something more? What is the distinction between expending 1700 calories versus 2200 calories per day with respect to the objective of shedding 5 pounds in next three months to improve the management of diabetes? Is there any abnormal behavior in body activity? For example, since the intake of asthma meds, if a Fitbit is showing the heartbeat at 100+ even during sleep - is that a serious enough condition requiring clinical consultation?

**(3) Self Management:** Self-management is the decisions and behaviors that a patient with chronic conditions engages in that impact their health. Generally, the impact that a patient may seek is getting back in line with the prescribed medical care plan or agreed upon health objective. Patients empowered with IoT-generated PGHD have a better sense of their health condition and make informed decisions about care as opposed to episodic clinical visit where patients are not aware of their state until diagnosed. An APH technology that intends to support self-management is expected to identify actionable information, such as increasing weight-bearing exercise or reducing consumption of energy-dense food. An APH technology can aid patients by providing alerts about potential triggers (e.g., high-pollen) or feedback on adherence (e.g., unexpected weight gain or not meeting activity targets), which they can use to stay on course. This improves the effective use of IoT for both data collection and relevant data/analysis/alert access by the patient. It can then provide alternatives to the patient to take steps to better adhere to the physician specified care plan to reduce adverse impact due to deviation from the plan or improve the outcome of the objective (e.g., an APH technology used to promote self-management for patients suffering from obesity can use IoTs such as activity monitoring and fluid consumption can measure increase activity level and targeted water intake if weight gain continues after use of oral steroid has ended).

**(4) Intervention:** Next step up in health management is clinical intervention which includes a change in the care plan prescribed by the clinician. An APH technology can use the data it gathers to help clinicians provide PGHD, environmental, and other data as well as corresponding analysis and interpretation to help evaluate and adjust a patient's clinical plan. The timely analysis of IoT data can yield insights for early intervention before a patient's situation deteriorates. In the case of kHealth Asthma that has developed an APH technology for managing asthma in children, the observed deterioration of asthma symptoms through PGHD may suggest change in medication or its dosage, develop trigger avoidance plans, etc. The IoT data collected from individuals can help the medical professionals to develop a personalized, patient-centric recommendation healthcare system with implicit feedback and support adherence to physician-prescribed protocols.

**(5) Disease Progression Tracking and Prediction**: Going beyond immediate and short-term management of health concerns, it would be highly rewarding both for the individual as well as public health if the longitudinal collection of personalized health data including PGHD and environmental data can facilitate tracking of how a disease is progressing, predict a significant change in health status, and take remedial actions. For example, for a prediabetic patient with an A1C score higher than 6, it would be highly valuable to be able to track progress toward a diabetic status (A1C ≥ 6.5), and predict the high probability of the patient becoming diabetic and requiring insulin treatment. For an overweight patient that is on long term steroid medication which leads to the number of adverse situation including higher energy intake, it would be important to track the worsening of weight and compute the probability of worsening asthma severity. A straightforward strategy would be for the clinician to periodically review the patient data and make an educated judgment on the disease progression. A more advanced strategy will be using the personalized health data and analyze it with the published clinical studies and longitudinal data collected with relevant cohort population.

The objective is to devise more proactive intervention and incorporate more of nonmedicinal solutions such as lifestyle changes that are often very effective but take a longer time to show the effect. Strategically, individual and public health will greatly benefit when what we learn from these strategies become evidence driven enhancement for broadly used clinical pathways and protocols.

**APH showcase and application scenario**

Knowledge-enabled Healthcare (kHealth) initiative at Kno.e.sis is an example of an APH framework to enhance decision making and improve health, fitness, and well-being (http://bit.ly/kAsthma). The early prototyping and testing involved kHealth - ADHF, a mobile app and a sensor kit targeted for reducing a readmission of patients with Active Decompensated Heart Failure (ADHF). The kHealth - ADHF involved continuous monitoring of observations using targeted questions driven by application specific (cardiovascular) knowledge and sensors including blood pressure, heart rate, and weighing scale that provided observational data via Bluetooth to the mobile application. The mobile health application also asked pertinent questions to the patient and analyzed all collected data and generated alerts. A follow-on application, kHealth - Asthma for better control of asthma in children extended physical data, with cyber and social data. Figure 1 shows an instance of kHealth-Asthma application on collecting multimodal data to monitor pediatric asthma, which is a multifactorial and multifaceted disease. We are running a trial with 200 asthma patient cohort collecting possibly the broadest and widest modality of data with an average 124 readings collected per day (2 tablet readings per day, 24 Fitbit readings per day, 2 Peak flow readings, 96 Foobot readings per day) for a duration of one, three and six months. This vast amount of multimodal and multisensory data poses a big data challenge (e.g. data variety, data integration) in comparison to other mHealth studies like Google Verily, IBM, and Swiss startup Docdok.health project and the Stanford wearable study which deals with fewer modality and lesser sample size. To address the aforementioned problem, kHealth supports contextual (condition specific) annotation, integration, and interpretation of sensor data using Semantic Sensor Network (SSN) ontology. Furthermore, kHealth supports contextualized actionable feature selection in PGHD to generate Smart Data using SSN and domain-specific knowledge sources. Utilization of Smart Data provides timely medical intervention and remediation measures. In a broad sense, a knowledge graph is a knowledge base that provides semantic annotation using its characteristic functionalities like fact extraction, named entity recognition, relationship identification, locale-specific information, event extraction, and intent identification to enrich information.

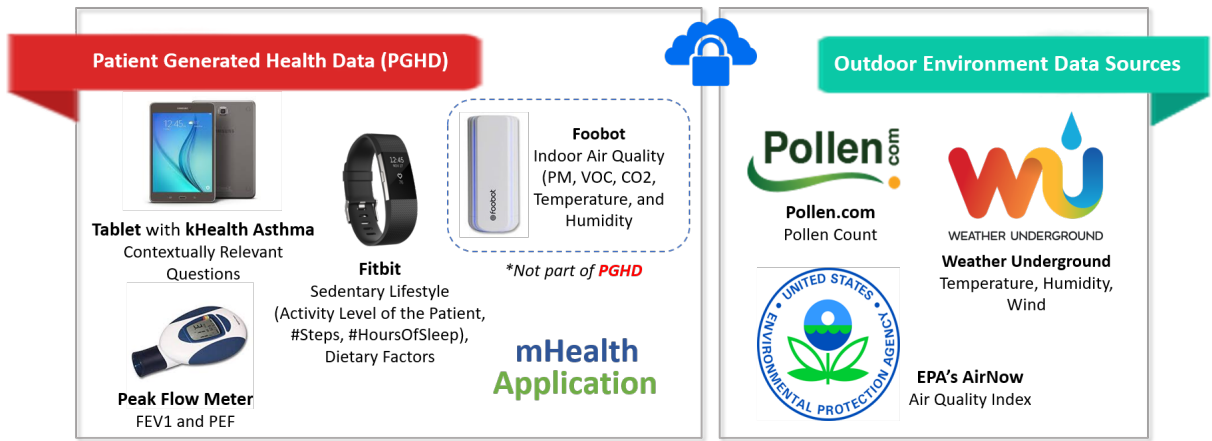

**Figure 1: kHealth, a mHealth application** that gathers PGHD data through contextually relevant questions (Tablet), sensors (Foobot, Peak Flow Meter), wearable (Fitbit), and from external data sources for contextual (e.g., location specific) environmental data.

We describe the kHealth APH approach with an example. Sara is a 10-year-old girl. With the help of our kHealth kit (Figure 1), she is able to monitor her daily activities, helping her and her clinician to intervene and update her treatment plan accordingly. Self-monitoring refers to the data collection using the mobile devices and sensors. Sara takes her mHealth application questionnaire twice a day, collects her daily activity level and sleep pattern using Fitbit [11] [http://bit.ly/1VdkW3I] and places an indoor air quality monitor to measure her indoor environment at home. Self-monitoring in many cases, will not be helpful if collected data are not acted upon. In general, we do not want a technology to make a clinical decision and change the care plan, but to enable adherence to a care plan specified by the patient's physician. Self-appraisal refers to the process of self-monitoring and self-reconciliation of the observed data. Self- management helps a patient to make better judgment or action within the scope of the care plan. For instance, you may observe that every time you have gone out and had allergic symptoms, the APH systems has found a strong correlation with high pollen. The intervention involves looping in the clinician for monitoring the severity level of patient and alter the care plan. Activities involved in the intervention may include the addition of new medication, changing the course of intake of medication or suggesting preventive medication considering the historical observations of the patient by involving the clinical and support services in care and health monitoring process. Assessment of post-intervention processes is crucial for re-classification of patient's disease. For instance, the collected evidences can help the physician to reclassify the asthma from mild persistent to moderate persistent and adjust the care plan with modification in the medication.

**Challenges in converting Big Data into Smart Data**

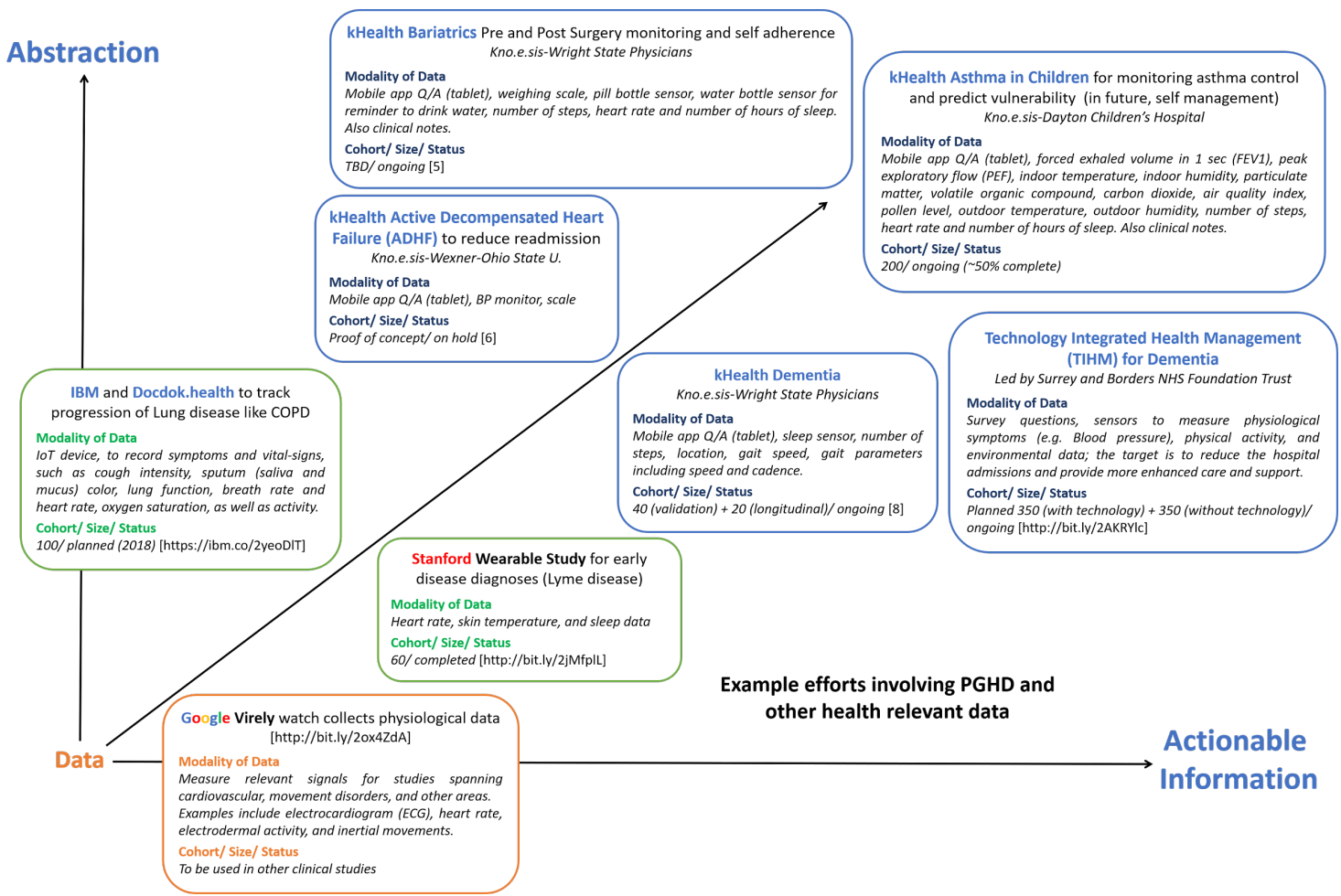

Figure 2: Example efforts involving PGHD and other health-relevant data. Converting PGHD data into Actionable Information

A variety of studies involving PGHD and other health-relevant data using a broad variety of IoT are ongoing for developing personalized digital care solutions for a variety of health related objectives, as shown in Figure 2. These systems need to deal with a host of data related challenges such as accessing, storing, querying, and managing large volumes of highly dynamic data, and systems related challenges such as interoperability and integration, security, privacy, trust, scalability, and reliability. [10] We characterized the ongoing efforts along two critical dimensions: abstractions (making the data meaningful and interpretable with respect to an individual's health) and actionable information (supporting decision making and actions informed from the data). These involve addressing challenges in data analysis including semantic data modeling, annotation, knowledge representation (e.g. modeling for constrained environments, complexity issues and time/location dependency of data), etc. While statistical analysis of the data collected helps one identify correlations, it is widely observed that a correlation does not necessarily imply a causation. Working with domain experts (that is, clinicians for the health applications) for understanding correlations between observations from different modalities is the key in associating meaning to the variations in observations that can then support derivation of causations. Another challenge is that the clinicians, health practitioners, and patients cannot keep up with an enormous amount of data being generated. Patients cannot interpret the data in the context of health conditions and objectives and clinicians do not have time to look at it. There is a dire need to convert the raw data into Smart Data. By making sense out of big data [http://j.mp/SmData], Smart Data provides value from harnessing the challenges posed by volume, velocity, variety, and veracity of big data, in-turn providing actionable information and improving decision-making process. Smart Data is focused on the actionable value achieved by human involvement in data creation, processing and consumption phases for improving the human experience [http://bit.ly/HumanExperience]. We propose the following evidence-based semantic perception approaches: (1) Contextualization, (2) Abstraction, and (3) Personalization.

**Contextualization** refers to data interpretation in terms of knowledge (context). PGHD consists of demographic and medical information from EMRs and time series data collected from various environmental sensors, physiological sensors, and public Web resources. Contextualization supports ranking of patient's diagnosis and patients similarity based on demographics and PGHD. It deals with these low-level fine-grain data covering various facets by determining the type and value of the data, and situates it in relation to other domain concepts, thus developing a meaningful interpretation of results. A large body of existing research on ontologies and Semantic Web techniques and technologies can be leveraged for this purpose [4]. However, relying solely on description logics or formal knowledge representation alone is often not sufficient to understand the complex nature of many health conditions. Probabilistic graph models from representing knowledge graphs, combined with machine learning and NLP on relevant data is an alternative some recent approaches have used.

**Abstraction** is a computational technique that maps and associates raw data to action-related abstractions, taking into account personal details but ignoring inessential differences to provide an integrated view of proper remediation measures. For example, high activity translates to different amounts of workout based on age, weight, current health, weather, and sport; or a low risk of heart problems depends on demographic and ancestry information, and food habits. In some cases, abstraction can be embedded on the device, for example, question answering system on mobile health applications as a way of indirectly supervised personalization of healthcare. However, one of the challenges is the need to formalize normalcy and detect an anomaly. Anomaly detection is non-trivial because the notion of normalcy itself is intrinsically dynamic, based on spatiotemporal and personal context, and requires personalization. It also requires uncovering various correlations among multi-modal data streams and discovering medically-relevant abstract interpretations and the factors that influence them. The challenge itself can be overcome if sufficient patient data can be obtained through large-scale clinical studies, followed by identification of correlations, and then analyzed and explained by those with domain knowledge and expertise to derive causations.

**Personalization** in health care refers to the determination of a treatment plan based on a patient's severity of the disease, the prevalence of triggers, and vulnerabilities with the use of past and current health data. For example, a low dosage SABA (Short-Acting Beta Agonists) may help someone to keep asthma symptoms in check in the fall season but it may not work for another patient who might have to resort to a higher dosage because of the higher severity of asthma and prevailing intensity of triggers in the spring season. IoT data provides an opportunity for personalization of future course of action and treatment plans by taking into account the contextual factors such as patient's health history, physical characteristics, environmental factors, activity, and lifestyle.

With contextualization, abstraction, and personalization in place, the next open problem is how to synthesize a personalized *vulnerability score* for a given medical condition or disease associated with a patient (a) with respect to a relevant health management objective to better capture and establish a control level, and (b) to quantify and express the effectiveness of remedial measures in a manner that is readily accessible to patient or clinician.

**Conclusion**
In terms of IoT and health, most of the current efforts are focused on data with collection and understanding what the data implies at a basic level. The data collected needs to be analyzed and validated with EMRs that capture patient's care objectives, plan, and self-reporting by patients. The key aspect is generating actionable information that will be acceptable, easy to use and can be integrated into clinical pathways be used by other systems, and be given at the right time with the right modality to the end beneficiaries (clinicians, patients) with appropriate information governance procedures and privacy/ security measures. The privacy issues in the healthcare data can be dealt by homomorphic encryption schemes, differential privacy, and data perturbation. Depending on the privacy level needed additional cryptographic services can be introduced in the framework.

Transitioning from a cohort-based treatment to a more personalized treatment, basic statistical computing with causality and machine learning algorithms will not suffice. There is a need to combine and integrate machine learning and data analytics with reasoning engines and knowledge bases, thus propelling us into the realm of augmented personalized health care management and well-being applications and services.

**Acknowledgements**

We thank Dr. Payam Barnaghi for his review and helpful suggestions. This work was partially supported by National Institutes of Health under the Grant Number: 1 R01 HD087132-01. The content of this paper is solely the responsibility of the authors and does not necessarily represent the official views of the National Institutes of Health.